\documentclass{mn2e}
\usepackage{epsf}
\usepackage{graphicx}

\def\msun{{\rm M_{\odot}}}
\def\rsg{{R_{\rm sg}}}
\def\msg{{M_{\rm sg}}}
\def\rw{{R_{\rm warp}}}
\def\dmm{{\Delta M_{\rm merger}}}
\def\dme{{\Delta M_{\rm episode}}}
\def\dau{{{\rm d}a\over{\rm d}M}_{\rm up}}
\def\dad{{{\rm d}a\over{\rm d}M}_{\rm down}}

\title[The Evolution of Black Hole Mass and Spin in AGN]
{The Evolution of Black Hole Mass and Spin in Active Galactic Nuclei}

\author[A.R. King, J.E. Pringle, J.A. Hofmann] {A.R. King$^1$,
J.E. Pringle$^{1,2}$ \& J.A. Hofmann$^3$\\ $^1$Theoretical Astrophysics
Group, University of Leicester, Leicester LE1 7RH\\ $^2$Institute of
Astronomy, University of Cambridge, Madingley Road, Cambridge CB3
0HA\\
$^3$Institut f\"ur Theoretische Physik und Astrophysik, Leibnizstrasse 15,
24118 Kiel, Germany}

\date{\today}

\volume{000}

\pagerange{\pageref{firstpage}--\pageref{lastpage}} \pubyear{2006}

\begin{document}

\label{firstpage}

\maketitle
\begin{abstract}
Observations show that the central black hole in galaxies has a mass
$M$ only $\sim 10^{-3}$ of the stellar bulge mass. Thus whatever
process grows the black hole also promotes star formation with far
higher efficiency. We interpret this in terms of the generic tendency
of AGN accretion discs to become self--gravitating outside some small
radius $R_{\rm sg} \sim 0.01 - 0.1$~pc from the black hole. We argue
that mergers consist of sequences of such episodes, each limited by
self--gravity to a mass $\dme \sim 10^{-3}M$, with angular momentum
characteristic of the small part of the accretion flow which formed
it. In this picture a major merger with $\dmm \sim M$ gives rise to a
long series of low--mass accretion disc episodes, all chaotically
oriented with respect to one another. Thus the angular momentum vector
oscillates randomly during the accretion process, on mass scales $\sim
10^3$ times smaller than the total mass accreted in a major merger
event.

We show that for essentially all AGN parameters, the disc produced by
any accretion episode of this type has lower angular momentum than the
hole, allowing stable co-- and counter--alignment of the discs through
the Lense--Thirring effect. A sequence of randomly--oriented accretion
episodes as envisaged above then produces accretion discs stably co--
or counter--aligned with the black hole spin with almost equal
frequency. Accretion from these discs very rapidly adjusts the hole's
spin parameter to average values $\bar a \sim 0.1-0.3$ (the precise
range depending slightly on the disc vertical viscosity coefficient
$\alpha_2$) from any initial conditions, but with significant
fluctuations ($\Delta a\sim \pm 0.2$) about these. We conclude (a) AGN
black holes should on average spin moderately, with the mean value
$\bar a$ decreasing slowly as the mass increases; (b) SMBH
coalescences leave little long--term effect on $\bar a$; (c) SNBH
coalescence products in general have modest recoil velocities, so that
there is little likelihood of their being ejected from the host
galaxy; (d) black holes can grow even from stellar masses to $\sim
5\times 10^9\, \msun$ at high redshift $z\sim 6$; (e) jets produced in
successive accretion episodes can have similar directions, but after
several episodes the jet direction deviates significantly. Rare
examples of massive holes with larger spin parameters could result from
prograde coalescences with SMBH of similar mass, and are most likely
to be found in giant ellipticals. We compare these results with
observation.

\end{abstract}


\section{Introduction}

Astronomers generally agree that the nuclei of most galaxies contain
supermassive black holes (SMBH), but as yet have little clear idea of
how they grow. Cosmological large--scale structure simulations
strongly suggest that galaxy mergers are the basic motor of growth,
and predict that a black hole typically acquires a mass $\dmm$ of
order its own mass $M$ in such mergers (e.g. di Matteo et al., 2005,
Li et al., 2007, and references therein). This is reasonable, as
growth rates of this order are needed in order to reach the hole
masses observed in the nearby universe. If one assumes further that
all the mass gain $\dmm$ carries the same sense of specific angular
momentum, the hole always spins rapidly up to Kerr parameters $a$
close to unity (e.g. Volonteri et al., 2005; Volonteri \& Rees,
2005). The resulting high radiative efficiency of accretion means that
the Eddington limit severely restricts the rate of black hole mass
growth, creating difficulties in understanding observations of high
masses at early cosmic times.

However it is sigmificant that $\dmm$ is only a very small part of the
total mass involved in a major galaxy merger, in line with the
observation that the black hole mass in nearby galaxies is typically
only $\sim 10^{-3}$ of the galaxy bulge mass. Moreover, current
cosmological simulations are entirely unable to resolve the
hydrodynamics of matter accretion on to the central black hole. The
mass inflow time through an accretion disc of 1 pc radius is already
$\ga 10^9$~yr (Shlosman, Begelman \& Frank, 1990), a factor $\ga 10$
greater than the timescale on which rapid hole growth occurs. The best
cosmological simulations resolve only lengthscales at least 100 times
larger. Inevitably the simulations have to resort to sub--resolution
recipes in trying to model the accretion process. For example it is
usual to assume accretion at the Bondi rate from some large radius,
whereas in reality the infalling gas must possess angular momentum
which will complicate things.

In this picture the very small mass $\dmm$ which accretes on to the
central black hole must have almost zero angular momentum about
it. Moreover, since the hole is growing its mass as rapidly as
possible, it must be close to its Eddington limit and thus feeding
back energy and momentum into its surroundings. Indeed simple models
of the resulting momentum--driven feedback on to the host galaxy
correctly reproduce the $M - \sigma$ relation between black hole mass
and velocity dispersion without any free parameter (King, 2003; 2005,
Murray et al., 2005).  Along with this, there must be copious star
formation from material close to the hole, i.e. at distances $\sim
0.01 - 0.1$~pc (see below) where the angular momentum neglected in
estimating the Bondi accretion rate becomes significant in supporting
the captured gas against gravity. This star formation also injects
energy and momentum into the gas near the hole. Hence although it is
very difficult to predict the gas flow near the black hole it seems
more likely to be chaotic than well--ordered, and rather inefficient
in growing the hole compared with forming stars, thus keeping the
latter a fairly constant fraction $\sim 10^{-3}$ of the stellar bulge
mass.

An obvious reason for the relative inefficiency of hole growth versus
star formation is that an accretion disc becomes self--gravitating at
radii $R$ where its mass exceeds $M_{\rm sg} \sim (H/R)M$, where $H$
is the disc scaleheight and $M$ the central (here black hole) mass
(e.g. Pringle, 1981, Frank et al. 2002). For AGN this means that discs
must be self--gravitating outside some radius $R_{\rm sg} \sim 0.01 -
0.1$~pc (Shlosman et al., 1990; Collin--Souffrin \& Dumont, 1990;
Hur\'e et al., 1994). Cooling times in these regions are so short that
self--gravity is likely to result in star formation (Shlosman \&
Begelman, 1989; Collin \& Zahn, 1999). As discussed by King \&
Pringle, 2007 (hereafter KP07) we expect that most of the gas
initially at radii $R > R_{\rm sg}$ is either turned into stars, or
expelled by those stars which do form, on a rapid (almost dynamical)
timescale, while the gas which is initially at radii $R < R_{\rm sg}$
forms a standard accretion disc, which slowly drains on to the black
hole and powers the AGN. KP07 show that for local, low--luminosity
AGN, fuelling by {\it well separated} episodes of this type explains
observational features such as the luminosity function for
moderate--mass black holes, and the presence and location of a ring of
young stars observed about the Galactic Centre.

This paper deals not with such low--luminosity AGN but instead with
high--luminosity events causing major black--hole mass growth. In
these cases our discussion suggests a picture in which, within an
event, the flow on to the hole is episodic, via a rapid succession of
accretion discs limited by self--gravity (as in the well--separated
episodes in low--luminosity AGN), and thus with masses $\dme \sim
\msg$. The number of such episodes is evidently $\sim
\dmm/\msg$. Given that the flow is constantly stirred up by energy and
momentum input, and that we are concerned only with the very small
fraction of the gas with almost zero angular momentum, it is
reasonable to expect the disc orientations to be essentially random.

We thus arrive at a picture of AGN accretion close to that suggested
by Sanders (1984) (see also Heckman et al., 2004, Greene \& Ho, 2007,
and KP07). This picture reproduces several characteristic features of
nearby AGN. In particular the random orientations of the accretion
episodes mean that the radio jets show no correlation with the
grand--design structure of the host, as observed for low--redshift AGN
(Kinney et al., 2000), and inferred at higher redshift by Sajina et al
(2007). Moreover the picture implies a luminosity function in broad
agreement with that observed for moderate mass central black holes
(KP07). Finally this picture provides a plausible explanation for the
ring(s) of young stars seen around the black hole in the centre of our
own Galaxy (Genzel et al., 2003; Lu et al., 2006).

Since observational properties such as the luminosity function of AGN
and quasars are sensitive only to the total mass increase $\Delta
M_{\rm merger}$, and not to its internal angular momentum budget, our
picture predicts the same values for these observables as found in
previous studies (e.g. Wyithe \& Loeb, 2003) and is therefore in broad
agreement with expected galaxy merger rates.

Clearly the proper hydrodynamic treatment of the kind of complex event
we envisage must await advances in computing power. However we can
gain some insight by simple arguments.

\section{SMBH spinup}

To understand how a supermassive black hole grows we have to know its
spin rate. This governs the accretion efficiency and thus, through the
Eddington limit, the maximum growth rate for the mass. The range here
is very large. Given an adequate mass supply, a low Kerr spin
parameter $a$ and consequent low accretion efficiency $\epsilon$
allows very much faster mass growth than a rapidly spinning hole with
$a \sim 1$ (cf. King \& Pringle, 2006).

Accretion on to a black hole affects its spin in two ways. First, it
torques the hole by adding matter carrying the specific angular
momentum of the innermost stable circular orbit (ISCO), and second,
the Lense--Thirring (LT) effect creates a viscously mediated torque
between the hole and any accretion disc not completely co-- or
counter--aligned with the hole spin. Gas within the disc follows
precessing orbits because of the LT effect, and the resulting viscous
dissipation tries to produce an axisymmetric situation by adjusting
the direction of the angular momentum vectors of both the hole spin
and the disc.

The LT effect is very important in AGN, as the accretion episodes all
initially have random directions with respect to the hole spin. As it
has a large lever arm, it operates more quickly than the accretion
torque, which cannot greatly change the hole spin until the hole has
significantly increased its mass. Accordingly, the LT effect
determines whether the disc and hole angular momenta are co-- or
counter--aligned long before the accretion torque has had a real
effect. Early studies of the LT effect (e.g. Scheuer \& Feiler, 1996)
concluded that it always produced co--alignment. Accordingly studies
of SMBH growth (e.g. Volonteri et al., 2005; Volonteri \& Rees, 2005)
argued that the main effect of the accretion torque was always to
produce spinup. However King et al. (2005; hereafter KLOP) showed on
very general grounds that an initially retrograde accretion disc with
total angular momentum $J_d$ would end up stably counter--aligned with
a hole of angular momentum $J_h$ provided that $J_d < 2J_h$ and the
initial angle $\theta$ between the angular momenta satisfied
\begin{equation}
\cos\theta < -{J_d\over 2J_h}. 
\label{align}
\end{equation}
(Throughout this paper $J_d, J_h$ and the Kerr parameter $a$ denote
the absolute values of these quantities.)

Thus Scheuer \& Feiler's (1996) conclusion that co--alignment always
occurs depended on their implicit assumption that the outer disc was
fixed, i.e. $J_d \gg J_h$. By contrast, KLOP's result implies that
counter--alignment occurs in a fraction
\begin{equation}
f = {1\over 2}\biggl(1 - {J_d\over 2J_h}\biggr)
\label{hemi}
\end{equation}
of accretion episodes.

In their analysis, KLOP considered a disc of fixed total angular
momentum $J_d$ misaligned with the hole. In reality any
misaligned disc must be warped at some radius $R_{\rm warp}$, in
general rather smaller than its outer edge. Only the part of the disc
around $\rw$ exerts an LT torque on the hole, so the correct
interpretation of $J_d$ in such cases is the total angular momentum
passing through $\rw$ during the alignment process. This interpretation
agrees with numerical simulations of the alignment process in such
cases (Lodato \& Pringle, 2006). Of course if the whole disc is warped
$J_d$ is simply the total angular momentum of the disc as before.

Equation (\ref{hemi}) shows that a random sequence of episodes with
$J_d \ll J_h$ must have $f \simeq 1/2$, so that co-- and counter--alignment
are almost equally common. Since the ISCO for retrograde accretion carries
higher specific angular momentum than for prograde accretion if the
Kerr $a$ parameter is significant (in the ratio 11:3 for maximal
spins) we see that spindown is more effective than spinup, ultimately
producing a slowly--spinning hole (Hughes \& Blandford, 2003 reach a
similar conclusion for the effect of repeated coalescences of spinning
holes, for the same reason; see also Colbert \& Wilson, 1995). King \&
Pringle (2006) used this argument to show how suitably random
accretion episodes could allow the growth of very large SMBH ($\sim
5\times 10^9\, \msun$) at high redshift $z \sim 6$ from even
stellar--mass seed black holes.

In this paper we show that the fact that self--gravity cuts off the
accretion discs in AGN feeding episodes implies $J_d < 2J_h$ in almost
all cases. A sequence of repeated episodes of this type with random
orientations therefore spins the hole down to low spin parameters, and
we find a mean value $\bar a \sim 0.2-0.3$.

\section{SMBH accretion episodes}

We first require a simple description of the properties of the
accretion disc during a feeding episode of the type described above. One
can regard an evolving disc as passing through a sequence of steady
states, so we follow KP07 in adopting the disc properties derived by
Collin--Souffrin \& Dumont (1990) in the context of AGN. These are
essentially the same as those derived by Shakura \& Sunyaev (1973) for
steady discs in close binaries. The disc surface density is

\begin{eqnarray}
\label{surfdensity}
\lefteqn{\Sigma = 7.5 \times 10^6\left( \frac{\alpha_1}{0.03} \right)^{-4/5}
\left(\frac{\epsilon}{0.1}\right)^{-3/5}}\nonumber \\
\lefteqn{\times \left( \frac{L}{0.1 L_E}
\right)^{3/5} M_8^{1/5} \left( \frac{R}{R_s} \right)^{-3/5} \,{\rm g
\, cm}^{-2}.}
\end{eqnarray}
Here $\alpha_1$ is the Shakura \& Sunyaev (1973) viscosity
parameter, $\epsilon$ is the accretion efficiency, with the
luminosity $L$ and accetion rate $\dot{M}$ related by
\begin{equation}
L = \epsilon \dot{M} c^2.
\end{equation}
Further
\begin{equation}
L_E = 1.4 \times 10^{46} M_8 \, {\rm erg \, s}^{-1},
\end{equation}
is the Eddington luminosity, $M_8$ is the black hole mass, $M$, in units of
$10^8$ M$_\odot$, $R$ is the radius and
\begin{equation}
R_s = 2.96 \times 10^{13} M_8 \, {\rm cm}
\end{equation}
is the Schwarzschild radius of the central black hole.

The mass of the disc $M(<R)$ inside radius $R$ is then
\begin{eqnarray}
\lefteqn{M(<R) = 2.94 \times 10^{34}\left( \frac{\alpha_1}{0.03} \right)^{-4/5}
\left(\frac{\epsilon}{0.1}\right)^{-3/5}} \nonumber \\
\lefteqn{\times \left( \frac{L}{0.1 L_E}
\right)^{3/5} M_8^{11/5} \left( \frac{R}{R_s} \right)^{7/5} \, {\rm g}.}
\end{eqnarray}

The disc semi--thickness $H$ obeys
\begin{eqnarray}
\lefteqn{\frac{H}{R} = 1.94 \times 10^{-3} \left( \frac{\alpha_1}{0.03}
\right)^{-1/10} \left(\frac{\epsilon}{0.1}\right)^{-1/5}} \nonumber \\
\lefteqn{\times\left(\frac{L}{0.1 L_E} \right)^{1/5} M_8^{-1/10}
\left( \frac{R}{R_s}\right)^{1/20}.}
\label{thickness}
\end{eqnarray}

The disc becomes self--gravitating when its mass reaches
\begin{equation}
M_d = M_{\rm sg} \simeq \frac{H}{R}M,
\label{sg}
\end{equation}
(e.g. Pringle, 1981) which occurs at radii $R \ge R_{\rm sg}$, where
\begin{eqnarray}
\label{gravradius}
\lefteqn{\frac{R_{\rm sg}}{R_s} = 1.13 \times 10^3 \left( \frac{\alpha_1}{0.03}
\right)^{14/27} \left(\frac{\epsilon}{0.1}\right)^{8/27}} \nonumber \\
\lefteqn{\times\left(
\frac{L}{0.1 L_E} \right)^{-8/27} M_8^{-26/27}}.
\end{eqnarray}
In line with our discussion in the Introduction we identify $M_{\rm
  sg}$ 
as the mass $\dme$ in an individual accretion episode.
The evolution timescale of the disc, and thus the characteristic duration of an
episode, is given by $\tau_{\rm sg} =
M_{\rm sg}/\dot{M}$, i.e.
\begin{eqnarray}
\label{gravtime}
\lefteqn{\tau_{\rm sg} 
= 1.12 \times 10^6 \left( \frac{\alpha}{0.03} \right)^{-2/27}
\left(\frac{\epsilon}{0.1}\right)^{22/27}} \nonumber \\ 
\lefteqn{\times\left( \frac{L}{0.1 L_E}
\right)^{-22/27} M_8^{-4/27} \,  {\rm yr}}.
\end{eqnarray} 
In line with our discussion above we expect a major merger event to
consist of a chaotic sequence of such episodes. These may overlap, or
be slightly separated in time. Our conclusions are independent of the
the time profile of the accretion, and depend only on the total mass
and angular momentum accreted. We differ from other treatments such as
those of Volonteri \& Rees (2005) and Volonteri et al. (2007) in not
following their assumption that all the mass in a given major merger
accretes at the same angle, even at the scale of the inner accretion
disc.

\section{The Evolution of Black Hole Mass and Spin}

We can now see how a sequence of accretion episodes limited by disc
self--gravity will affect the black hole mass and spin. We compare the
angular momentum $J_d$ of the warped disc with the angular momentum
$J_h$ of the hole. The interpretation of $J_d$ as the total angular
momentum passing through $\rw$ shows that
\begin{equation}
J_d \la \msg(GM\rw)^{1/2}
\label{jd}
\end{equation}
(i.e. the total accreted disc mass $\dme = \msg$ multiplied by the
disc specific angular momentum at $\rw$ ), provided that $\rw$ is less
than the outer disc radius $R_{\rm sg}$. In this latter case we
replace $\rw$ with $R_{\rm sg}$. The warp radius is given by (cf
KLOP)
\begin{eqnarray}
\lefteqn{\frac{\rw}{R_s} = 990 \left(\frac{\alpha_1}{0.03}\right)^{1/8}
\left(\frac{\alpha_2}{0.03}\right)^{-5/8} a^{5/8}}
\nonumber \\
\lefteqn{\ \ \ \ \ \ \times \left(\frac{\epsilon}{0.1}\right)^{1/4}
\left(\frac{L}{0.1L_E}\right)^{-1/4} M_8^{1/8}.}
\label{warp}
\end{eqnarray}
Here $R_s = 2GM/c^2$ is the Schwarzschild radius and $a$ the
dimensionless Kerr spin parameter, with $\alpha_2$ the dimensionless
viscosity coefficient relevant to vertical disc motions, and we note
that Lodato \& Pringle (2007) have recently argued that $\alpha_2$
never exceeds a value $\sim 1$ in a warped disc.

If the condition $\rw < R_{\rm sg}$ fails we instead take
\begin{equation}
J_d = M_{\rm sg}(GMR_{\rm sg})^{1/2},
\label{jsg}
\end{equation}
which implies that $J_d$ can never be larger than given by (\ref{jd}).

Given the hole angular momentum
\begin{equation}
J_h = 2^{-1/2}Ma(GMR_s)^{1/2},
\end{equation}
we find from (\ref{jd})
\begin{eqnarray}
\frac{J_d}{2 J_h} \le 0.061 \, a^{-11/16} \left( \frac{\alpha_1}{0.03}
\right) ^{-5/432} \left( \frac{\alpha_2}{0.03} \right)^{-5/16} \nonumber \\ 
   \times \left( \frac{\epsilon}{0.1} \right)^{-13/216}
\left( \frac{L}{0.1 L_E} \right)^{13/216} {M_8}^{-37/432} \; .
\label{ratio}
\end{eqnarray}
\begin{figure}
  \centerline{\epsfxsize9cm \epsfbox{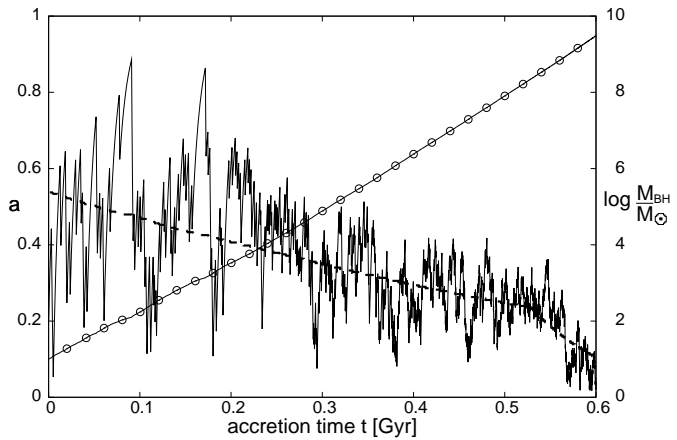}}
  \caption{Numerical simulation of black hole mass $M$ (circles) and
  spin parameter $a$ (solid curve) evolution versus accretion time
  (i.e. neglecting epochs when no mass accretes). The hole is assumed
  to accrete always at its current Eddington rate, through a sequence
  of disc accretion episodes with disc angular momentum ${\bf J}_d$
  randomly oriented with respect to that of the hole, ${\bf J}_h$. (In
  reality these episodes can be separated in time.) The value of $J_d$
  is limited by eqn (\ref{jd}), or by eqn (\ref{jsg}) if the warp
  radius $R_{\rm warp}$ exceeds the self--gravity radius $R_{\rm
  sg}$. The dashed curve shows the expected means spin parameter $\bar
  a$ (see eqns \ref{equil}, \ref{av}), and bends sharply at lower
  right, reflecting the change from eqn (\ref{jd}) to eqn
  (\ref{jsg}). In this simulation the hole has initial mass $M_0 =
  10\msun$ and spin parameter $a_0= 0.2$, but $a$ very rapidly
  converges towards the mean value $\bar a$ (cf eqn \ref{av}) whatever
  its initial value. We take viscosity parameters $\alpha_1 = \alpha_2
  = 0.03$. Note that the hole mass grows more slowly at epochs when
  $a$ is large, and more rapidly when it is small.}
\label{afig}
\end{figure}
\begin{figure}
  \centerline{\epsfxsize9cm \epsfbox{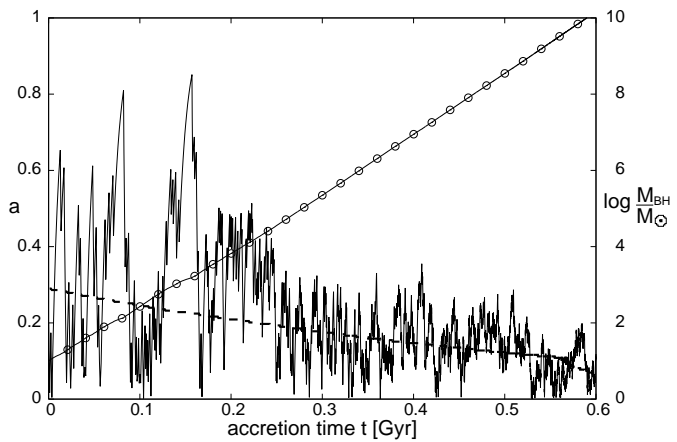}}
  \caption{As for Figure \ref{afig}, but with $\alpha_2 = 1$ The same
  random sequence of accretion episode orientations is used in order
  to highlight the effect of changing $\alpha_2$.}
\label{a3fig}
\end{figure}

We see that the ratio $J_d/2J_h$ determining the tendency of the disc
to co-- or counter--align under LT torques is always $<1$ unless the
spin parameter $a$ is small itself, effectively independently of all
other parameters except for a weak dependence on $\alpha_2$. With
$J_d/2J_h <1$ retrograde accretion episodes stably counter--align with
the hole angular momentum, spinning it down.  Since spindown is more
efficient than spinup, the general effect of a sequence of
randomly--oriented accretion episodes as envisaged in the Introduction
is to decrease $a$. Thus one might imagine that the end result is a
value of $a$ oscillating around zero, and both King \& Pringle (2006)
and Volonteri et al (2007) assume this. However as $J_h$ decreases for a
given $J_d$, the probability $p({\rm spindown})$ of accreting in a
stably retrograde fashion (given by $f$ in (\ref{hemi})) approaches
zero. Evidently $J_h$ must on average decrease only to the point where
the expected spinup in prograde accretion is equal to the expected
spindown in retrograde accretion, i.e.
\begin{equation}
p({\rm spindown})\dad = p({\rm spinup})\dau
\end{equation}
or equivalently
\begin{equation}
p({\rm spindown}) = {1\over 2}\biggl(1 - {J_d\over 2J_h}\biggr) =
{\dad\over \dad + \dau}.
\label{equil}
\end{equation}
One has to use the equations of Bardeen (1970) to evaluate the
derivatives in this equation, which then defines the mean value $\bar a$.

\subsection{Numerical Simulations}
We simulate the effect of repeated accretion episodes of the type
dicussed here as follows. We assume that each episode has mass $\msg$
and accretes at the Eddington rate appropriate to the current black
hole mass $M$. The total angular momentum $J_d$ of each episode is
given by the recipe (\ref{jd}), replacing $\rw$ by $\rsg$ if $\rw >
\rsg$.  The direction of ${\bf J}_d$ is chosen at random from an
isotropic distribution. The disc and hole are assumed to co-- or
counter--align according to the criterion (\ref{align}). In contrast
to Volonteri et al (2007) we do not assume that all the mass accreting
in a major merger has the same orientation of angular momentum, but
only the mass $\msg$ contained within each self--gravitating disc
episode, allowing each episode to be randomly oriented.

Figures \ref{afig}, \ref{a3fig} show the results of two such
simulations. The figures omit the epochs when the hole is not
accreting, so the horizontal axes measure the accretion time rather
than the total elapsed time. Since the accretion is always at the
current Eddington rate, the accretion time is the shortest possible
time for the hole to acquire its mass through accretion.

The two simulations differ only in the value of the `vertical'
viscosity coefficient $\alpha_2$, which is 0.03 in Fig. \ref{afig} and
1 in Fig \ref{a3fig}. The two simulations use the same random sequence
of accretion episode orientations in order to highlight the effect of
changing $\alpha_2$.
The main difference is that the mean value $\bar
a$ is somewhat lower in the second case (see below).

\begin{figure}
  \begin{center}
    \begin{tabular}{cc}
      \hspace{-0.8cm}
      \resizebox{45mm}{!}{\includegraphics{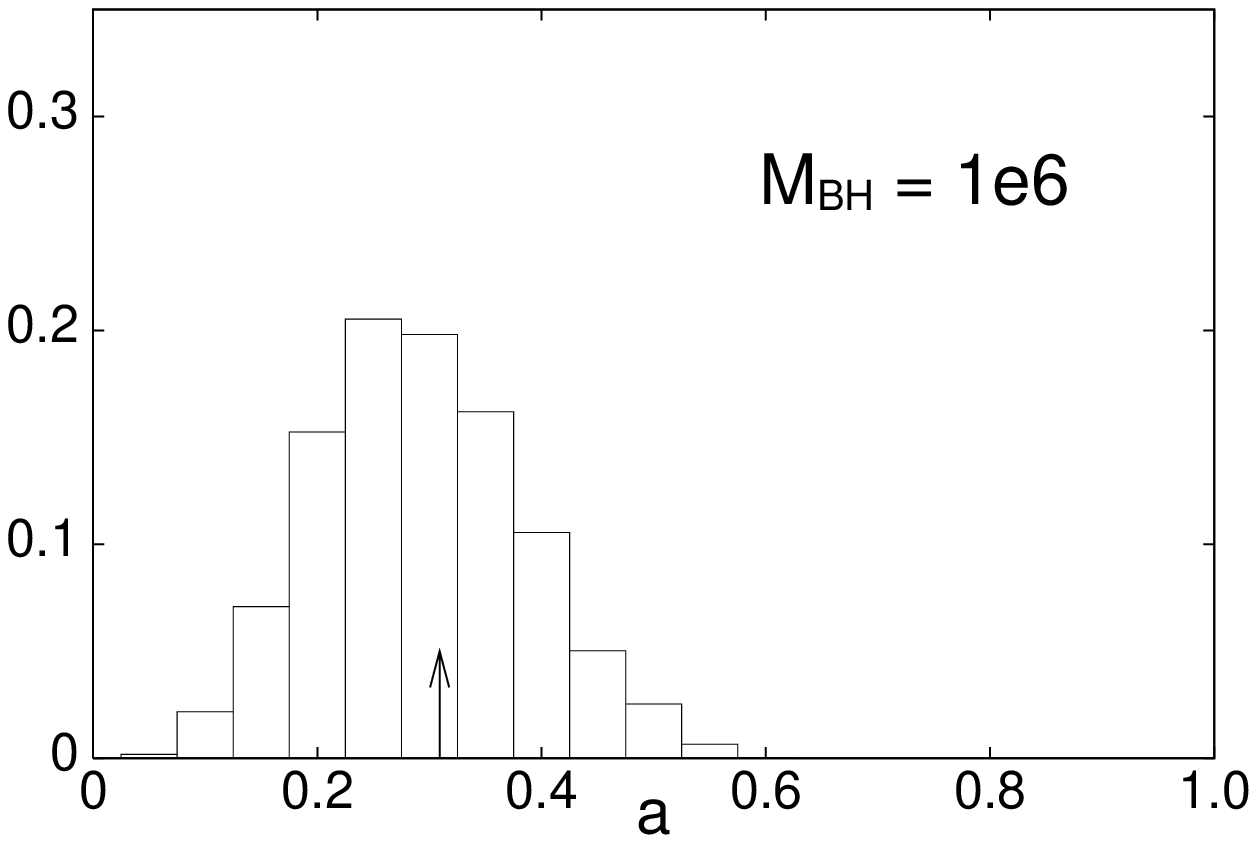}} 
&
      \hspace{-0.5cm} \resizebox{45mm}{!}{\includegraphics{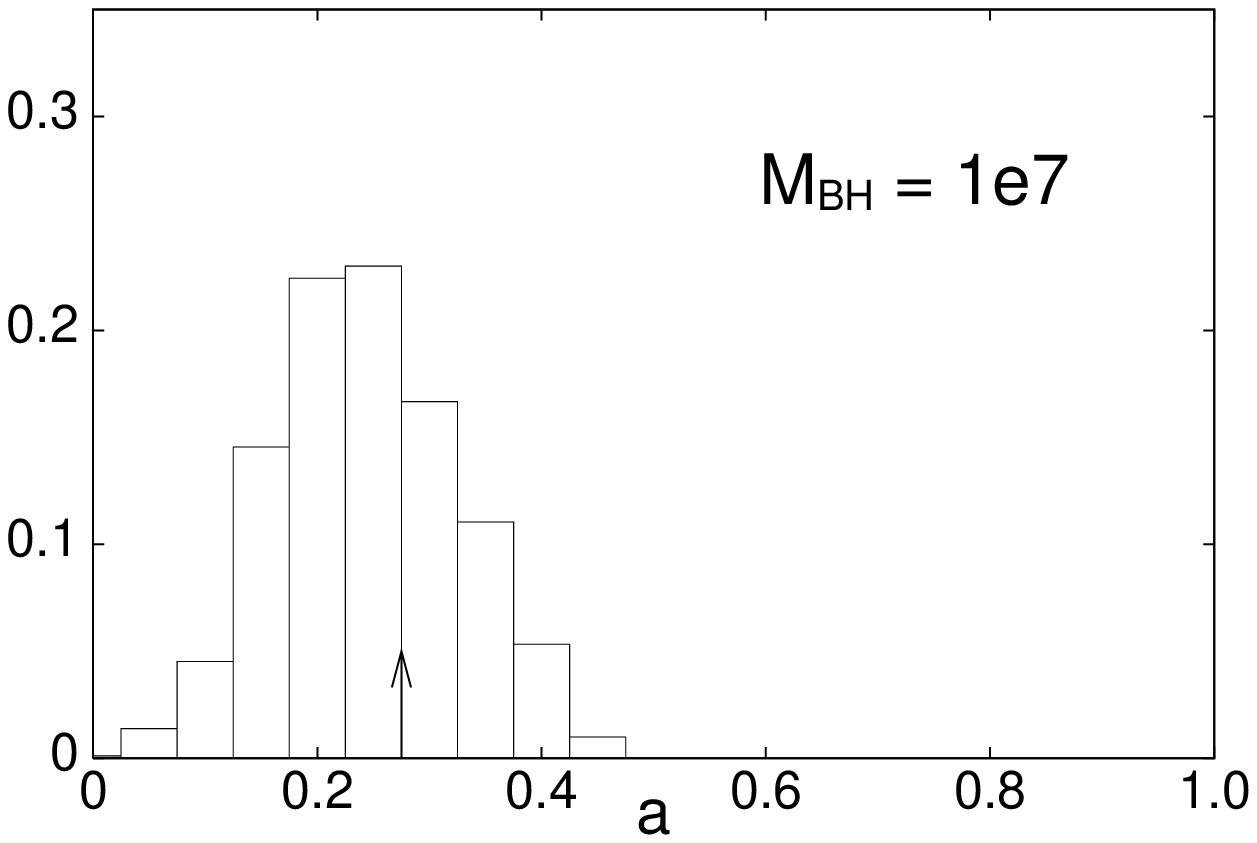}} 
\\
      \hspace{-0.8cm} \resizebox{45mm}{!}{\includegraphics{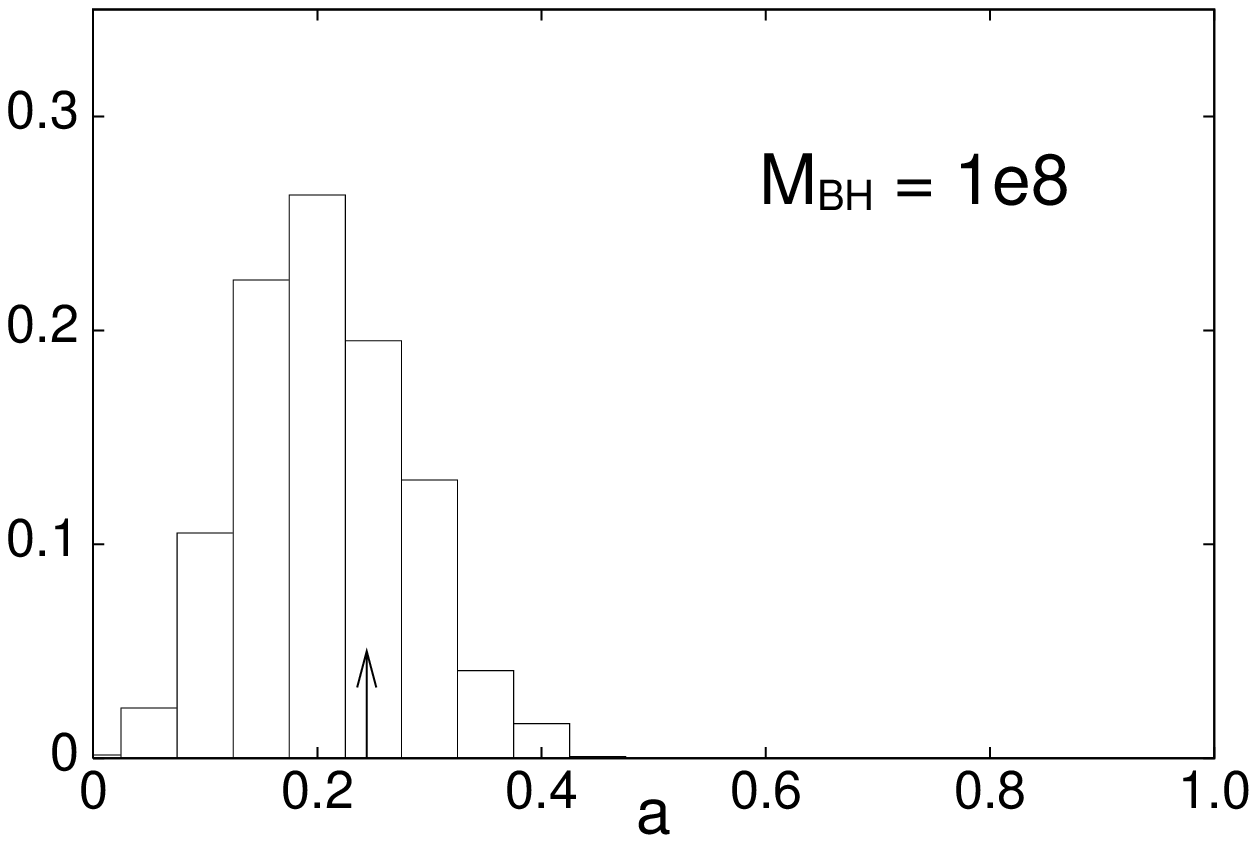}} 
&
      \hspace{-0.5cm} \resizebox{45mm}{!}{\includegraphics{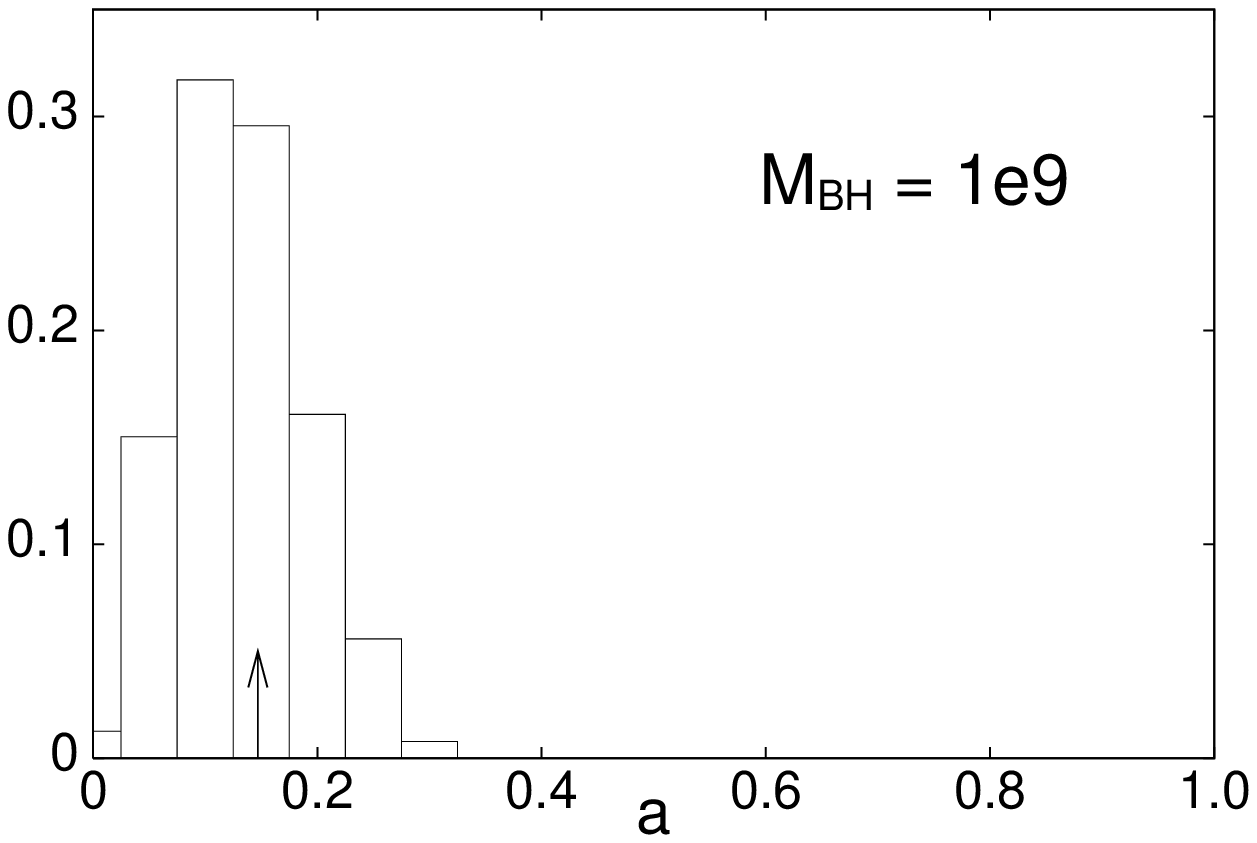}} 
\\
    \end{tabular}
    \caption{Distribution of $a$ around the mean $\bar a$ (arrow) at
    masses $M/\msun = 10^6, 10^7, 10^8, 10^9$, for the case $\alpha_2
    = 0.03$.}
  \end{center}
\label{hist_03}
\end{figure}

\begin{figure}
  \begin{center}
    \begin{tabular}{cc}
      \hspace{-0.8cm} \resizebox{45mm}{!}{\includegraphics{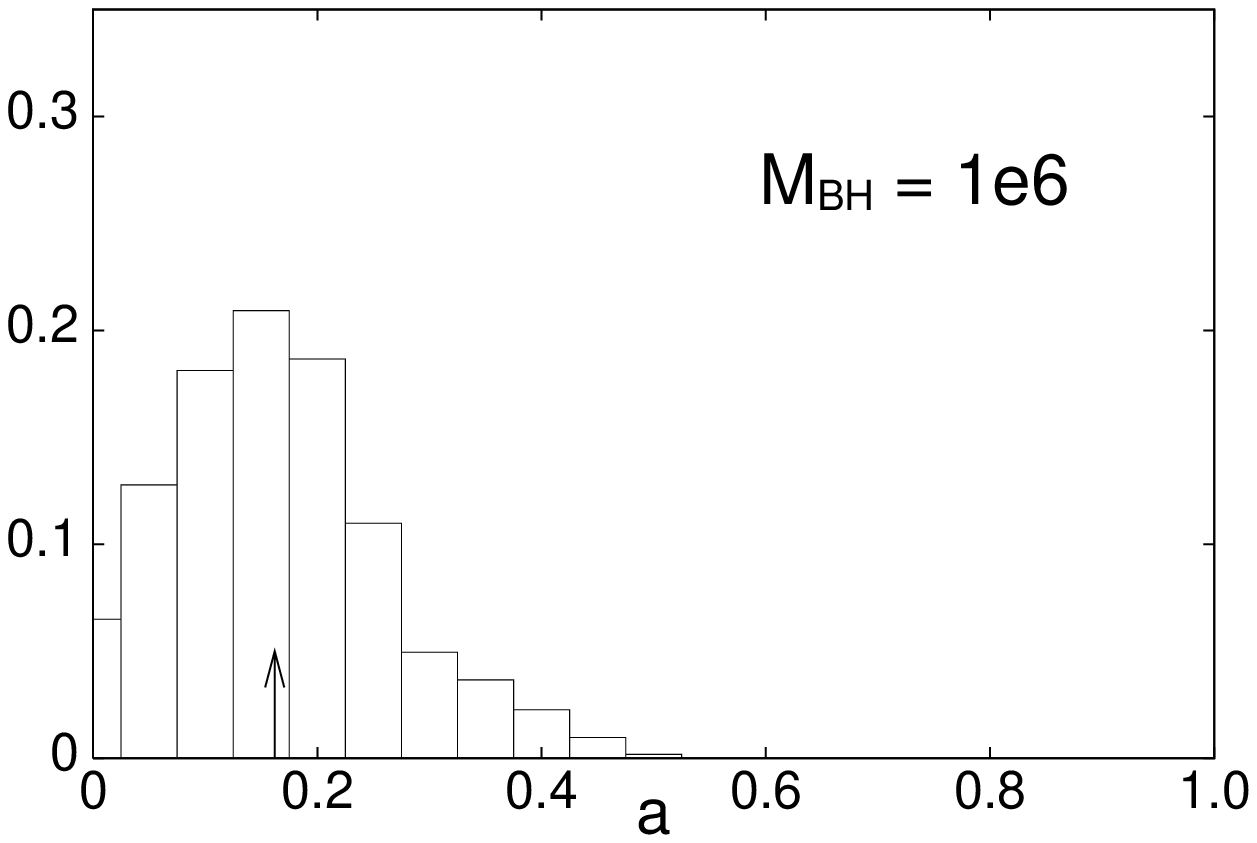}} &
      \hspace{-0.5cm} \resizebox{45mm}{!}{\includegraphics{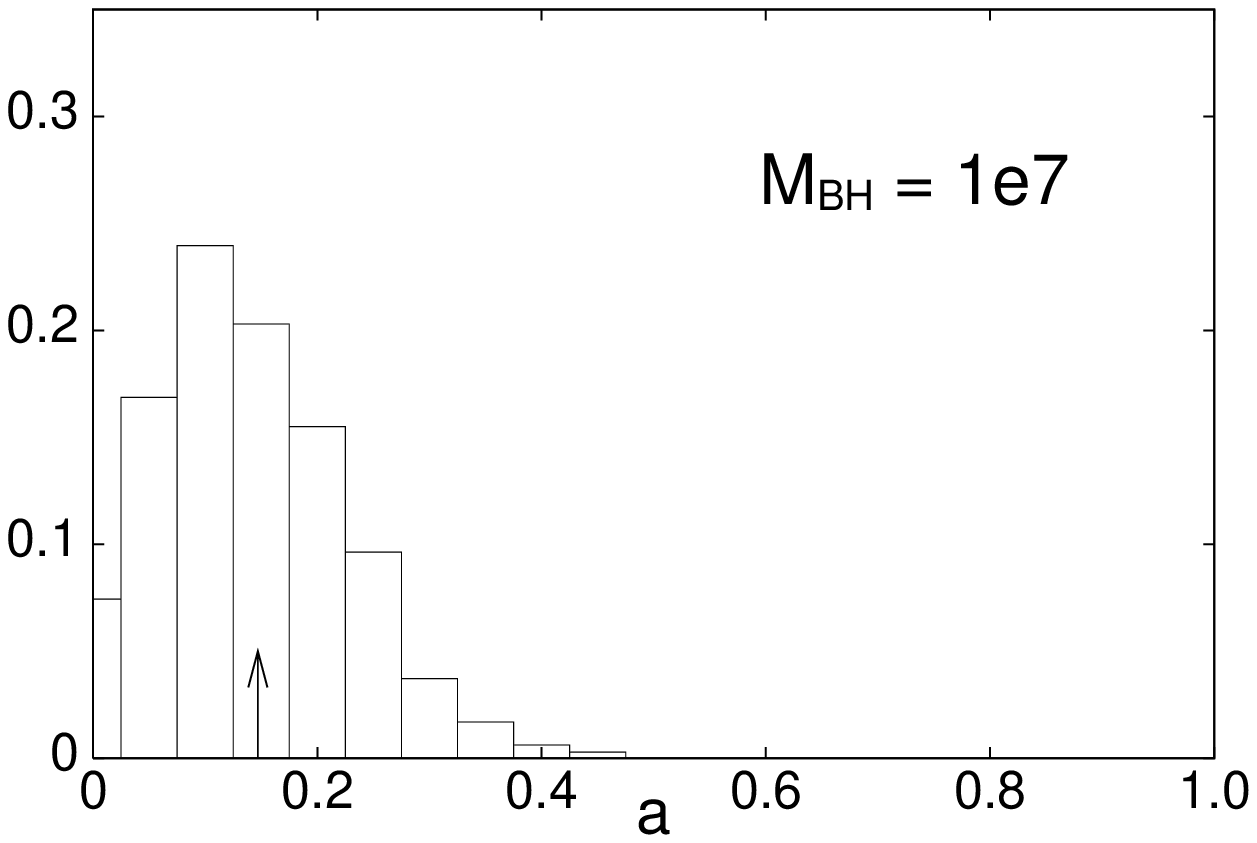}} \\
      \hspace{-0.8cm} \resizebox{45mm}{!}{\includegraphics{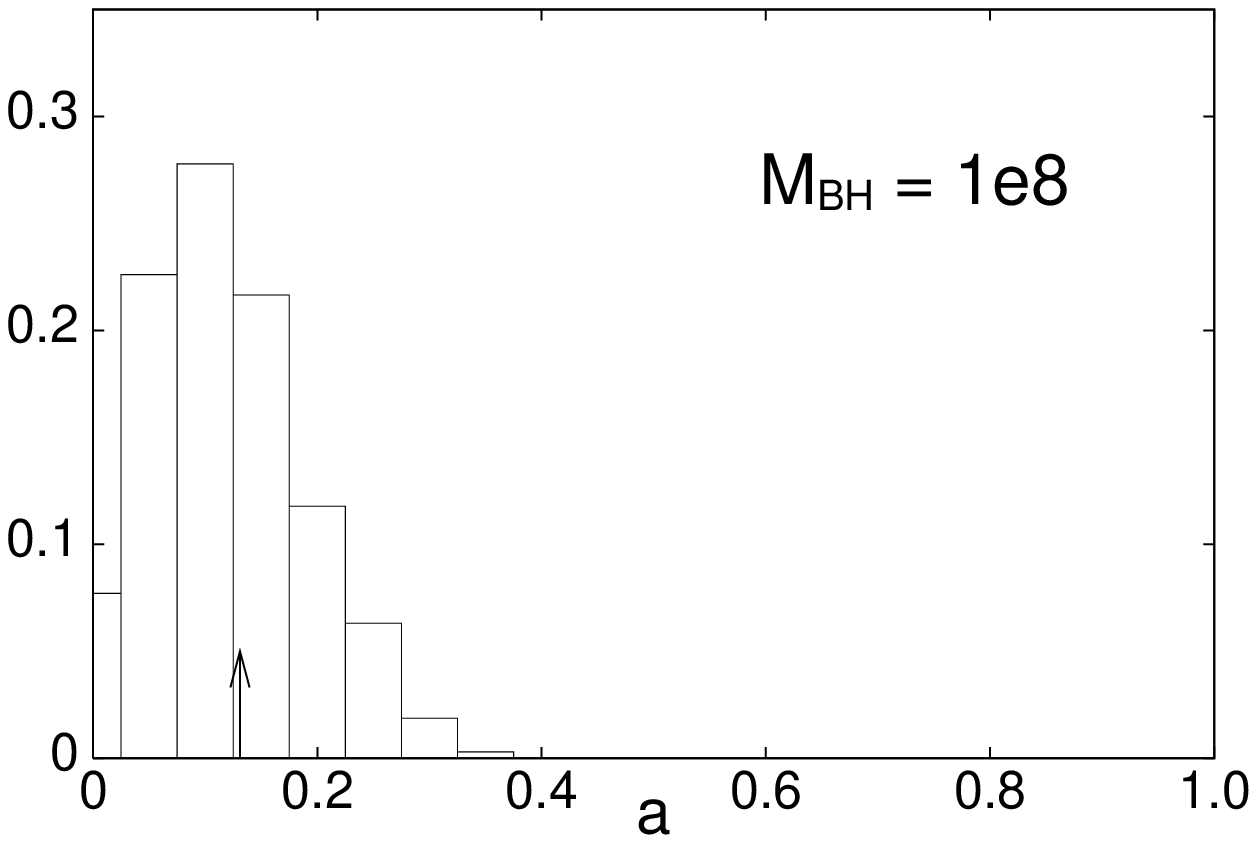}} &
      \hspace{-0.5cm} \resizebox{45mm}{!}{\includegraphics{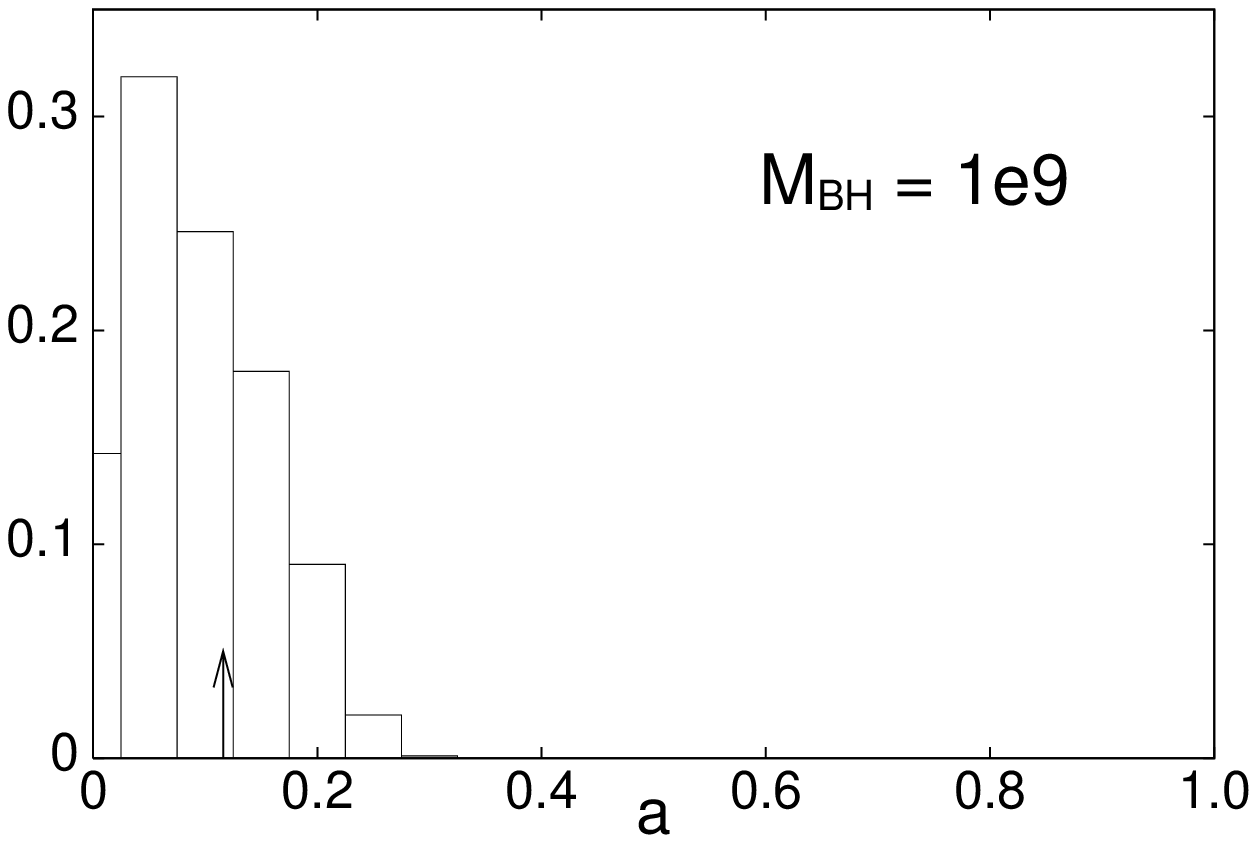}} \\
    \end{tabular}
    \caption{As Figure 3, but for $\alpha_2 = 1.0$.}
  \end{center}
\label{hist_1}
\end{figure}

As one can see, the main features inferred above do appear. Although
the value of $a$ fluctuates widely, its mean $\bar a$ does indeed
quickly settle to the value predicted by (\ref{equil}) in each
case. We can fit this as
\begin{eqnarray}
\lefteqn{\bar a \simeq AM_8^{-0.048},\ \  R_{\rm warp} < R_{\rm sg}} 
\nonumber \\
\lefteqn{\ \   \simeq 0.3M_8^{-0.3},\ \ R_{\rm warp} > R_{\rm sg}}
\label{av}
\end{eqnarray}
where $A = 0.246$ (for $\alpha_2 = 0.03$), $0.13$ (for $\alpha_2 = 1$)
is fixed by continuity at the changeover between the two power--law
regimes in each case. In Fig. \ref{afig}, this occurs at $M_8 \simeq
2$, which in the case with higher vertical viscosity the changeover is
only at a mass of almost $10^{10}\msun$. We see that for holes of
masses $\sim 10^6 - 10^9\msun$ we expect $\bar a \sim 0.3 - 0.2$ in
the first case and $\bar a \sim 0.2 - 0.1$ in the second, with
excursions $\Delta a \sim \pm 0.2$ about these means. Figures
\ref{hist_03}, \ref{hist_1} show the normalized distributions of $a$
as a function of mass in the case $\alpha_2 = 1$.


\section{SMBH Coalescences}

We have so far discussed the evolution of SMBH mass and spin via
accretion. Figures \ref{afig}, \ref{a3fig} show that the Kerr
parameter $a$ always has a strong tendency to move towards the mean
value $\bar a$ from any initial value as the hole mass doubles through
accretion. Thus although a coalescence of two SMBH with comparable
masses would presumably cause a discontinuous departure from the mean
accretion--driven trend of Figs. \ref{afig}, \ref{a3fig}, accretion
would drive it back towards the mean trend as it doubles the mass of the
coalesced hole. In other words coalescences have little long--term
effect on $\bar a$, although they can strongly affect the spin for a
relatively short time (see the end of the Discussion).

\section{SMBH Recoil Velocities}

The relatively modest spin rates we find mean that SMBH coalescences
even with similar masses (i.e. in major mergers) are likely to produce
fairly small gravitational radiation recoil velocities $\la
200$~km\,s$^{-1}$ (Gonz\'alez et al., 2006). These are well below the
escape velocity from the merged host. This agrees with the
observational evidence that most massive galaxies do have nuclear
SMBH. We note that Bogdanovi\'c et al. (2007) reach similar
conclusions, but for entirely orthogonal reasons. Specifically, they
argue that the effect of gas accretion is to cause any SMBH to align
with the angular momentum of the accretion flow. They further argue
that this flow is of such large scale and well--defined angular
momentum that in any incipient SMBH coalescence both holes have their
spins aligned with the flow and thus with each other. It is known that
co--aligned spins reduce the recoil velocity to modest values $\la
200$~km\,s$^{-1}$, similar to those for non--spinning holes as we
found above.

Our arguments differ from these in three ways. First, we have argued
above that self--gravity severely limits the angular momentum of any
accretion flow so that $J_d < 2J_h$, so that co-- and
counter--alignment occur almost equally often, rather than the hole
and disc always co--aligning as asserted by Bogdanovi\'c et
al. Second, we have also argued that the accretion flow on to an SMBH
cannot have as large a scale as claimed by Bogdanovi\'c et al., who
suggest a linear size of $10^5R_s \sim 1$~pc, because the viscous
timescale from such radii exceeds a Hubble time. Finally Bogdanovi\'c
et al.'s picture predicts that the SMBH spin and the accretion flow
are both aligned with the large--scale structure of the galaxy. One
would thus expect any jet structure to show a similar alignment
othogonal to the galaxy disc apparently feeding it. However this
disagrees with the observational evidence discussed by Kinney et al
(2000; see also Nagar \& Wilson 1999): in a sample of nearby ($z \leq
0.031$) Seyfert galaxies the direction of the jet from the central
black hole is unrelated to the orientation of the disc of the host
galaxy. As mentioned above Sajina et al. (2007) have recently inferred
a similar result at higher redshift.

\section{Maximum Mass Growth Rate for SMBH}

In an earlier paper (King \& Pringle 2006) we considered the problem
of growing the largest SMBH masses $\sim 5 \times 10^9\, \msun$ at
cosmological redshift $z \simeq 6$, i.e. only $\sim 10^9$~yr after the
Big Bang. We concluded that growth to these masses from even stellar
initial values was possible provided that the spin rate, and thus the
radiation efficiency of mass accretion, was kept sufficiently low ($a
\la 0.5$). From the discussion of the last section we see that this
condition is always satisfied by accretion from randomly--oriented
accretion episodes. The main limit on black hole mass growth is
therefore the mass supply rather than the Eddington limit. Figure \ref{afig}
shows this explicitly, giving $M$ as a function of the accretion
time.

\section{SMBH spin directions}

We have so far considered the magnitude $J_h$ of the black--hole
angular momentum vector ${\bf J}_h$ rather than its direction. The
latter may be important, as the central part of the accretion disc
always has its axis parallel or antiparallel to it, and this is
probably the direction of any jet produced by the system. Observations
of various types support this conclusion. Schmitt et al (2003) show
that the extended [O III] emission in a sample of nearby Seyfert
galaxies is well aligned with the radio, and it is usual to assume
that the orientation of the [O III] emission is governed by the
geometry of the inner torus, of typical radius 0.1 -- 1.0 pc
(e.g. Antonucci, 1993). Similarly, Verdoes--Kleijn and de Zeeuw (2005)
and Sparks et al (2000) find evidence that dust discs tend to be
perpendicular to radio and optical jets respectively.

As the torques between the disc and the hole are all internal, the
outcome of an accretion event is that the hole spin vector aligns
along the total a.m. vector ${\bf J}_t = {\bf J}_h + {\bf J}_d$ (cf
KLOP). We can therefore track the direction of ${\bf J}_h$ over
time. From eqn (\ref{ratio}) we see that $J_d \ll J_h$ for $a \sim 1$,
and even for $a$ near the mean value (\ref{av}) we have $J_d/J_h \la
0.26, 0.16$ for the two cases $\alpha_2 = 0.03, 1$. By simple vector
addition we have that in such an accretion episode the vector ${\bf
J}_h$ deviates from its original direction by an angle $\psi$ given by
\begin{equation}
\sin\psi = {J_d\over J_h}\sin\theta \la 0.26, 16
\end{equation}
for any original misalignment angle $\theta$, so that $\psi \la
15^{\circ}, 9^{\circ}$ for $\alpha_2 = 0.03, 1$. Thus successive
accretion episodes have a tendency to produce jets in similar
directions, even more so if $a$ happens to be larger than
average. This may explain why some `double--double' radio sources are
seen with apparently near--common projected axes for successive jet
events. Kharb et al. (2006, Figs. 1, 5) give a beautiful example of
this. Ultimately the effect of even a few episodes is that the spin
axis loses all memory of its original direction. There is clearly no
correlation at all with structures in the host galaxy. Coalescences
have similar disorienting effects.

\section{Discussion}

We have argued that accretion on to supermassive black holes in AGN
occurs through events with total mass $\dmm \sim M$, consisting of
long sequences of randomly oriented disc accretion episodes with
masses $\dme \sim 10^{-3}M$. The mass of these disc episodes is
strongly constrained by the generic tendency of such discs to become
self--gravitating and form stars outside quite small radii $\sim 0.01
- 0.1$~pc. The self--gravity constraint also limits the disc angular
momentum. This is always less than the maximum that the hole's mass
would allow it to have. A hole with significant spin therefore stably
co-- and counter--aligns initially prograde and retrograde discs
respectively, producing a net spindown until its angular momentum is
reduced to a relatively modest value, with fluctuations $\Delta a \sim
\pm 0.2$ about the average $\bar a$ given by (\ref{av}). We note from
(\ref{av}) that $\bar a$ itself decreases slowly as the black hole
mass decreases.

We can compare these results with those for coalescences of SMBH as a
consequence of the mergers of their host galaxies. Hughes and
Blandford (2003) show that the average secular evolution of black hole
spin under coalescences is
\begin{equation}
a = a_0\biggl({M_0\over M}\biggr)^{2.4}
\end{equation}
where $a_0$ is the original spin parameter of a hole within initial
mass $M_0$. Thus coalescences doubling the mass of a hole on average
reduce $a$ by a factor $\sim 5.3$. This is an even faster decline of
spin than given by assuming fixed angular momentum but growing mass,
which would produce $a \propto M^{-2}$, reflecting the fact that
retrograde coalescences are more effective in reducing spin than
prograde ones are in growing it. Evidently coalescences can reduce $a$
quite quickly to zero. However we have seen that accretion very
rapidly pushes $a$ back towards the mean value $\bar a$ (cf eqns
\ref{equil}, \ref{av}) from any starting point. Hence on average
coalescences have little net effect on the mean value of $a$.  We
conclude that repeated accretion episodes and coalescences on average
drive the black holes in AGN towards fairly low values of $a$, with
statistical fluctuations around them of order $\Delta a = \pm 0.2$.

\begin{figure}
  \centerline{\epsfxsize9cm \epsfbox{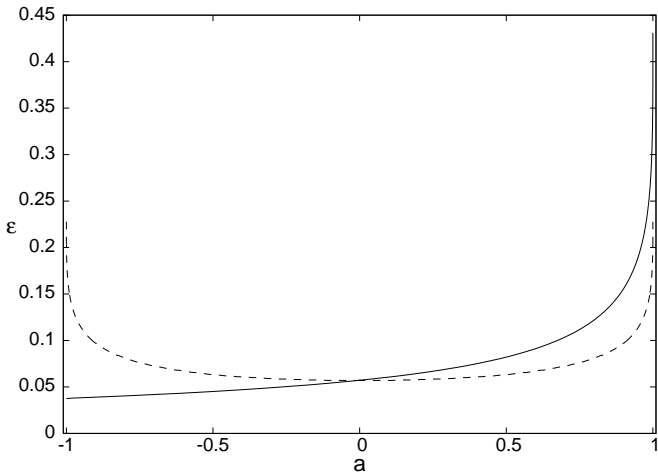}}
  \caption{Accretion efficiency $\epsilon$ versus spin parameter $a$
  (solid curve). Negative values of $a$ denote retrograde accretion. The
  dashed curve shows the effective $\epsilon=\epsilon_{\pm}$ given by assuming
  that retrograde and prograde accretion are equally probable. This is
  the efficiency relevant to the Soltan argument.}
\label{epsilon}
\end{figure}

Given this result, we should consider the observational evidence
constraining values of $a$ for SMBH. Much of this is derived from the
Soltan (1982) argument that the totality of background light in the
low--redshift Universe is compatible with black--hole growth with
radiative efficiency $\epsilon \simeq 0.1$ (e.g. Wang et al., 2006;
Treister \& Urry, 2006; Hopkins, Richard \& Hernquist, 2007), which
formally requires $a \simeq 0.67$. However $\epsilon$ is such a slow
function of $a$ except near $a=1$ (see Figure \ref{epsilon}) that this
conclusion is vulnerable to the systematic effects inherent in trying
to estimate the efficiency from observation. Evidently we have to find
$\epsilon \simeq 0.1$ from observation to an accuracy of a factor $\la
2$ in order to exclude {\it any} value of $a$ other than those very near
unity. Since retrograde and prograde accretion appear equally
probable, the effective efficiency for the Soltan argument is actually
the symmetrized curve
\begin{equation}
\epsilon_{\pm} = {1\over 2}[\epsilon(a)+\epsilon(-a)] 
\label{epspm}
\end{equation}
in Figure (\ref{epsilon}), whose vertical range is now compressed to
$0.057 - 0.230$.  In this case only a very accurate estimate of
$\epsilon$ can give real information about $a$ at all.

Streblyanska et al. (2005) suggest a different method of estimating
the typical spin $a$ for SMBH. They find that the composite
X--ray spectrum of a sample of AGN shows evidence for a broad iron
line. They fit lines produced with assumed emissivity and line centre
energy from both non--rotating and rotating black holes, and argue
that the outer radius of the emission region for the non--rotating
hole is unreasonably small in the case of the Type I AGN in their
sample, although the statistical fit is marginally preferred over a
rotating hole. The best--fitting rotating hole for this group has an
inner emission radius of $3GM/c^2$, which requires $a > 0.78$. Since
holes with such values of $a$ would automatically have higher
efficiencies and thus presumably stronger iron lines, it is difficult
to know what fraction of the SMBH in the sample actually have such
fairly large spin parameters (note that, as discussed below, our
arguments only refer to the statistical mean $\bar a$, and individual
SMBH can differ significantly from this).

We remark as a general caveat that the search for any effect
depending on $a$, or the interpretation of a particular observational
effect as being due to $a$, is likely to produce a bias towards $a
\sim 1$. This effect is particularly marked because of the very high
efficiency of prograde accretion for $a \simeq 1$ (see Fig
\ref{epsilon}). We note that Brenneman \& Reynolds (2006) estimate $a
= 0.989^{+0.009}_{-0.002}$ in the specific case of the broad iron line
in MCG-06-30-15, remarkably close to the maximum allowed value.

Of course since our treatment is statistical it does not rule out
larger spin values in individual cases. The most important way of
producing large values of $a$ in individual cases is in coalescences
of two SMBH. Hughes \& Blandford (2003) show that this can produce
significant $a$ for prograde coalescences of two black holes with
similar mass. This suggests that giant ellipticals offer a promising
site for significant values of $a$. Another conceivable way of
producing exceptions to our general conclusions is that we have
assumed accretion from thin discs, in which cooling is efficient, so
that the total disc mass is limited to $\sim H/R$ times that of the
hole. In some situations it may happen that cooling is inefficient and
our conclusions do not hold. More work is needed here, although such
cases are likely to be rare. Finally one might argue that in some
cases an accretion event with mass $\gg M_{\rm sg}$ may contrive to
retain some memory of an overall angular momentum, despite the fact
that it must have a near--radial orbit within the host galaxy in order
to accrete at all (cf. Kendall et al., 2003: recall that accretion
within a Hubble time requires disc radii $\la 1$~pc). In this case
individual episodes no longer have uncorrelated angular momenta as we
assumed. This is effectively what is assumed by Volonteri \& Rees
(2005), and Volonteri et al (2007) for {\it all} episodes.

\section{Acknowledgments} 

ARK acknowledges a Royal Society Wolfson Research Merit Award. JH
acknowledges a scholarship from the Deutscher Akademischer Austausch
Dienst (DAAD) and partial support from the Deutsche
Forschungsgesellschaft (DFG) through grant SFB 439.

\label{lastpage}

\end{document}